# TRANSCENDING THE LEAST SQUARES

Fyodor V. Tkachov

Institute for Nuclear Research
of Russian Academy of Sciences
Moscow, 117312, Russia

The method of quasi-optimal weights provides a comprehensive, asymptotically optimal, transparent and flexible alternative to the least squares method. The optimality holds for a general non-linear, non-gaussian case.

> *"La méthode qui me paroit la plus simple et la plus générale consiste à rendre minimum la somme des quarrés des erreurs, ... et que j'appelle méthode des moindres quarrés."*
>
> Adrien-Marie Legendre, 1806

> *"So the last shall be first, and the first last…"*
>
> Matthew 20:16

**1. INTRODUCTION.** The least squares method [1] is a staple of experimental sciences.[a] However, physicists involved with the ever more delicate and expensive experiments — and with the computing power at their disposal exponentiating by Moore's law — are increasingly impatient with a sub-optimality of the venerable method in the non-linear, non-gaussian situations, beyond the limits of the Gauss-Markov theorem (see e.g. sec. 7.4.4. in [6]), that they have to deal with. Two examples are: (1) the neutrino mass measurement experiment [10] observing Poisson-distributed events, that has been providing context and a test bed for the methods described in the present Letter, and that will be referred to for the purposes of illustration; (2) the very high precision muon anomalous magnetic moment measurements where a non-trivial special case of the method we advocate was discovered [12] and put to use [18].

The concept of counting *weighted* events was advanced by Karl Pearson in [2] and is known as the method of moments. In Pearson's original treatment, power weights $x^n$ were used, motivated partly by the need to simplify his formidable computations (involving data on 1000 crabs), partly by the then popular mathematical problem of moments (restoring a function from its integrals with $x^n$), whence the terminology. A generalization to arbitrary continuous weights is obvious, and we henceforth use — following refs. [12] and [18] — the term weights as the most suggestive one in place of the somewhat obscure "moments" or quantum-theoretic "observables".

After the arrival of the maximal likelihood (ML) method in Fisher's research [3], Pearson's trick was stigmatized as non-optimal[b] — despite the obvious analytical advantages it offered, and for that reason alone deserving a scrutiny rather then the role of a mathematical warm-up in textbooks before the *complicated, therefore interesting* issues are addressed. The stigma is manifest in how its presentation always has to be justified by its historic priority — followed by the non-optimality mantra. From the Particle Data Group summaries of statistical methods, its mention has been dropped altogether for a number of years already.

Yet the analytical transparency of the method remains a temptation.

The development of functional analysis that occurred after Kolmogorov postulated his axioms for probability [4] — in particular, the emergence of the concept of generalized function interpreted as a linear functional on suitably chosen "test functions" [5] — supplied theorists with some very powerful calculational techniques [13].[c] From the functional-analytic point of view, a measure (probability distribution) is *defined* as a linear functional on arbitrary continuous test functions $\varphi$ — which for a non-singular probability distribution corresponds to its integrals with continuous weights $\varphi$. This highlights a basic nature of the concept of weighting, and induces one to regard Pearson's method as a foundational one. With one's mind thus focused, a meditation brings forth a simple question: *Which weight yields an estimate with lowest variance?*

---

[a] Coincidentally, it was first made public exactly two centuries ago — a fact that should have been a cause for symposia, but is, surprisingly, not.

[b] The rivalry between the two men must have contributed to this.

[c] Note also how test functions are curiously similar to acceptance functions of elementary detector cells in modern high energy physics detectors (about $10^5$ cells of several types and sizes).

Even if the question was too simple for mathematicians to bother with, practical needs compelled a number of physicists to arrive, in specific ways, at specific answers — special cases of Eq.(6) below (see [11] and further references to this line of research in [16]). Curiously, the formulae in those papers are similar to those in section 8.2.2 of [6] — yet the section is explicitly devoted to a demonstration that the method of moments is less efficient than the ML method. Also, the fact that the differential maximum condition of the ML method can be formally derived from the method of moments with a specially chosen weight (cf. Eq.(8)) was recorded, in small print, in a meticulous treatise [9] — but remained disconnected from a traditional cursory discussion, complete with the mantra, of Pearson's method; the small print paragraph stopped short of realizing the significance of the fact.

It is simply incredible how experts — pre-conditioned by the non-optimality mantra — would bump into the key formulae yet fail to see the whole picture.

A different and complex special case of an optimal weight was independently worked out in [12] and chosen for data processing in a significant measurement [18].

In short, the concept of counting weighted events was virtually begging to be given a most serious consideration.

Refs. [14], [16], [19] (stemming from the earlier findings of powerful calculational methods based on generalized functions [13]) promulgated the functional-analytic approach to the problems of statistics as advantageous over the traditional one. The advantages are due, in the final respect, to the resulting true algebraization of the subject (because the term $\sigma$-algebra is, really, a masquerade — exacerbated by a constructive deficiency of the notion of a function on arbitrary sets). Ref. [14] tried to shed a systematic light on Pearson's unfairly neglected trick and develop it into a general method:

## 2. THE METHOD OF QUASI-OPTIMAL WEIGTHS.

Given a sample of events $\{\mathbf{P}_i\}_i$ governed by a probability distribution $\pi(\mathbf{P})$ that depends on a parameter $\theta$ whose exact value $\theta^*$ is unknown, one estimates $\theta^*$ by choosing a weight $\varphi(\mathbf{P})$ and equating the theoretical mean,

$$\langle\varphi\rangle_{th} = \int d\mathbf{P}\,\pi(\theta)\,\varphi(\mathbf{P}) \equiv h(\theta), \qquad (1)$$

assumed to be a calculable function of $\theta$, to the corresponding experimental value

$$\langle\varphi\rangle_{\exp} = \frac{1}{N}\sum_i \varphi(\mathbf{P}_i), \qquad (2)$$

and to solve the resulting equation

$$h(\theta) = \langle\varphi\rangle_{\exp} \qquad (3)$$

to obtain an estimate $\theta_{\exp}$ for the unknown value $\theta^*$:

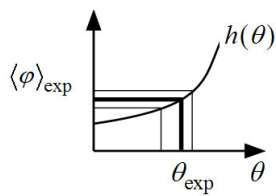

$$(4)$$

The procedure is perfectly transparent and yields estimates with nice properties [6], [9].

**NB** The procedure remains meaningful even when the classical assumption of the asymptotic normality of the distribution of $\theta_{\exp}$ does not hold for whatever reason: with modern computing power, it may be possible to perform a brute-force study of the distribution of $\theta_{\exp}$.

But if (theoretical) $\mathrm{Var}\,\varphi$ exists, then, for large $N$,

$$N\,\mathrm{Var}\,\theta_{\exp} = \frac{\mathrm{Var}\,\varphi}{H^2}, \qquad H = \frac{\partial h(\theta_{\exp})}{\partial\theta}. \qquad (5)$$

Ref. [14] showed that under technical assumptions similar to those of the ML method, the minimum of $\mathrm{Var}\,\theta_{\exp}$ is located in the space of $\varphi$ at

$$\varphi_{\mathrm{opt}}(\mathbf{P}) = \frac{\partial\ln\pi(\mathbf{P})}{\partial\theta}. \qquad (6)$$

The corresponding calculation would be a piece of cake already for Euler or Bernoullis: one deals here with a ratio of two quadratic forms of $\varphi$, so the intricacies of defining functional derivatives etc., are essentially irrelevant.

Moreover, the deviations from the value of $\mathrm{Var}\,\theta$ at (6) are non-negative and quadratic in $\varphi - \varphi_{\mathrm{opt}}$ (see ref. [14] for details), and the resulting minimum happens to exactly correspond to the Fisher-Fréchet-Rao-Cramér boundary (according to [9], the inequality is essentially known to Fisher in the 20's). The quadraticity is of a critical importance as it ensures that even imperfect approximations to $\varphi_{\mathrm{opt}}$ would yield practically optimal results.

One would start with an arbitrarily chosen weight $\varphi_{(0)}$ and obtain a valid (if non-optimal) estimate $\theta_{\exp(1)}$ — complete with an error estimate. One would then use this to construct the corresponding $\varphi_{\mathrm{opt}(1)}$ and use the latter to obtain an estimate $\theta_{\exp(2)}$ together with an improved variance. As to the initial weight $\varphi_{(0)}$, one can take it to be the optimal weight $\varphi_{\mathrm{opt}(0)}$ corresponding to some $\theta_0$ (because one should be able to construct $\varphi_{\mathrm{opt}}$ for any $\theta$ anyway). The successive weights $\varphi_{\mathrm{opt}(0)}, \varphi_{\mathrm{opt}(1)}$ etc. are optimal for the values of $\theta$ that are not necessarily equal to the unknown exact $\theta^*$, whence the prefix "quasi."

Each estimate in the resulting chain

$$\theta_0 \to \theta_{\exp(1)} \to \theta_{\exp(2)} \to \ldots \qquad (7)$$

asymptotically (for large $N$) converges to the true value of $\theta$ for any initial $\theta_0$, so the issue of convergence of iterations is theoretically (i.e. for large $N$) non-existent. The purpose of iterations is, basically, to reduce the variance of the estimate. In the context of [10], a few iterations often prove to suffice (thanks to the quadraticity).

**NB** The described iterations together with those implicit in the solution of (3) correspond to the optimizations of the conventional methods.

**NB** Note a simple and useful identity that holds under the obvious regularity assumptions:

$$\langle\varphi_{\mathrm{opt}}\rangle = 0. \qquad (8)$$

**NB** The resulting estimate will not be affected (unless the non-linearities at a finite $N$ are large on the scale of the variances) if one drops a **P**-independent factor or addendum from (6). However, dropping addenda (that usually originate from normalization factors of probability densities) spoils the useful (although not strictly necessary) property (8).

### 3. THE MULTI-PARAMETER CASE.[d]

With several parameters $\theta_\alpha$ to be estimated, one needs an equal number of weights $\varphi^{(\alpha)}(\mathbf{P})$ and the corresponding functions $h^{(\alpha)}(\theta)$. Eq. (17) becomes a system of equations (one for each weight) to determine the vector $\theta_{exp}$.

**NB** As in the single-distribution case, the vector weight $\varphi^{(\alpha)}(\mathbf{P})$ is defined up to a **P**-independent matrix factor or vector addendum.

What is the quantity to minimize in the multi-parameter case? The issue is important because unlike the one-parameter case where one deals with a single scalar $\text{Var}\,\theta$, here one deals with the covariance matrix $\text{Covar}\,\theta \equiv \Theta$. Also, the issue of selecting one among several possible solutions when solving the system (3), is more pronounced with many parameters, and a proper criterion is of value.

Let us follow the functional-analytic line of reasoning. Let $\theta^*$ be the unknown exact value of the vector variable $\theta$, and let $\theta_{exp}$ be the solution of the system (17) for some fixed set of weights $\varphi$. $\langle\varphi\rangle_{exp}$ is a random vector constructed from $\{\mathbf{P}_i\}_i$ — denote its probability distribution as $\tilde{\pi}\left(\langle\varphi\rangle_{exp}\right)$.

The first, sine qua non assumption is that the latter distribution converges, for large $N$, to the $\delta$-function $\delta\left(\langle\varphi\rangle_{exp} - h(\theta^*)\right)$. This assumption already makes Pearson's method meaningful as it allows one to obtain an estimate $\theta_{exp}$ by solving the system (3).

To pursue optimality, one assumes existence of second moments, and makes a more detailed statement that $\tilde{\pi}\left(\langle\varphi\rangle_{exp}\right)$ approaches the limiting $\delta$-function when $N \to \infty$ via a multi-dimensional normal distribution centered at $h(\theta^*)$:

$$\tilde{\pi}\left(\langle\varphi\rangle_{exp}\right) \to \left(\frac{N}{2\pi}\right)^{\frac{k}{2}} |\det\Phi|^{-1/2}$$

$$\times \exp\left[-\frac{N}{2}\left(\langle\varphi\rangle_{exp} - h(\theta^*)\right)^{\mathrm{T}} \Phi^{-1}\left(\langle\varphi\rangle_{exp} - h(\theta^*)\right)\right]$$

$$\to \delta\left(\langle\varphi\rangle_{exp} - h(\theta^*)\right), \qquad (9)$$

where $k$ is the number of parameters and $\Phi$ is the theoretical covariance matrix for the weights. Then the asymptotic probability distribution of $\theta_{exp}$ is described by a similar expression,

$$\left(\frac{1}{2\pi}\right)^{\frac{k}{2}} \times \frac{\exp\left[-\frac{1}{2}\left(\theta_{exp} - \theta^*\right)^{\mathrm{T}} \Theta_{exp}^{-1}\left(\theta_{exp} - \theta^*\right)\right]}{\left|\det\Theta_{exp}\right|^{1/2}} \qquad (10)$$

with the covariance matrix (cf. Eq. (5))

$$N\,\Theta_{exp} = H^{-1}\Phi\left(H^{-1}\right)^{\mathrm{T}}, \qquad (11)$$

where $H$ is the matrix of the tangent mapping for $h(\theta)$:

$$H_{\alpha,\beta}(\theta) = \frac{\partial h^{(\alpha)}(\theta)}{\partial \theta_\beta}. \qquad (12)$$

The error hyperellipsoid for $\theta_{exp}$ for a given confidence level has the hypervolume $\sim |\det\Theta_{exp}|^{1/2}$. The missing numerical coefficient is determined by the confidence level chosen, whereas the exact shape and orientation of the hyperellipsoid (a thin disk or a cigar, etc., variously oriented) — and, most importantly, its hypervolume depend on the choice of weights.

So, the determinant

$$N^k\left|\det\Theta_{exp}\right| = \frac{|\det\Phi|}{\left(\det H\right)^2} \qquad (13)$$

is the first candidate to consider for the role of a single quantity to minimize in order to find the optimal weights.

The choice of determinant also respects physical dimensionalities of parameters.

**NB** One might wish to minimize the volume of projection of the hyperellipsoid to the subspace of some parameters deemed more interesting than others. (In the context of [10], the interesting parameter would be the neutrino mass; the other parameters describe uncertainties in the knowledge of the experimental setup.) This would require a more complicated analysis beyond the scope of this Letter. However, the standard results summarized in [9] concerning the multidimensional FFRC inequality, seem to indicate that nothing of interest can be found in this direction.

The requirement of minimizing the determinant (13) leads to a reasoning that follows the one-parameter case considered in [14] in a straightforward fashion: one evaluates the variational derivatives of (13) with respect to $\varphi^{(\alpha)}$ and equates them to zero to obtain a set of equations to determine the point of minimum, etc. To get past the determinants, one employs the following algebraic identity for derivatives:

$$\left(\det A\right)' = \left(\det A\right) \times \mathrm{Tr}\left[A' \times A^{-1}\right]. \qquad (14)$$

With the determinants out of the way, the remaining calculations are completely straightforward if a little cumbersome. One eventually arrives at the same result as if each parameter were considered separately (as stated in [14]):

$$\varphi_{opt}^{(\alpha)}(\mathbf{P}) = \frac{\partial \ln \pi(\mathbf{P})}{\partial \theta_\alpha}. \qquad (15)$$

Thus equipped, we are ready to tackle our main problem.

---

[d] This section properly belongs to [14].

## 4. THE MULTI-DISTRIBUTION CASE.

First consider the case of one unknown parameter. Let each event $\mathbf{P}_i$ in the sample $\{\mathbf{P}_i\}_i$ be governed by its own distribution $\pi_i(\mathbf{P}_i)$. Various $i$ correspond to experimental conditions controlled or recorded by the experimenter (Fisher's independent variables). All $\pi_i(\mathbf{P}_i)$ are supposed to be known modulo the unknown value $\theta^*$ of a common parameter $\theta$. For example, in the studies of rare decays in [10], one encounters Poisson distributions with the means depending via known formulae on the unknown neutrino mass. Another class of examples is considered below in sec. 6.

Some $\pi_i$ may be equal — the formalism must still work correctly. The case when all of them are equal, is the basic situation already considered (several equally distributed events). This will provide an important guidance.

With independent $\mathbf{P}_i$, one can regard the sample $\{\mathbf{P}_i\}_i$ as a cumulative event whose probability density is $\pi(\{\mathbf{P}_i\}_i) = \Pi_i \pi_i(\mathbf{P}_i)$. Then the optimal weight is immediately obtained as follows:

$$N\,\varphi_{\text{opt}}\big(\{\mathbf{P}_i\}_i\big) = \frac{\partial \ln \pi\big(\{\mathbf{P}_i\}_i\big)}{\partial \theta}$$
$$= \sum_i \frac{\partial \ln \pi_i(\mathbf{P}_i)}{\partial \theta} = \sum_i \varphi_{\text{opt},i}(\mathbf{P}_i). \qquad (16)$$

The inconsequential factor $N$ is introduced to faciliate comparison with the single-distribution case.

The weight (16) is already enough for a direct generalization of the previously described method with all its nice properties preserved: one estimates $\theta^*$ by solving the equation

$$N h(\theta) \equiv \sum_i h_i(\theta) \equiv \sum_i \int d\mathbf{P}\, \pi_i(\mathbf{P})\,\varphi_i(\mathbf{P}) = \sum_i \varphi_i(\mathbf{P}_i),\; (17)$$

with $\varphi_i$ taken from (16) with some initial $\theta_{(0)}$, etc.

**NB** If one drops a $\mathbf{P}$-independent factor from the definition of the weight, it must be the same for all $\varphi_i$. One can drop a different $\mathbf{P}$-independent addendum from each $\varphi_i$ (usually such terms correspond to normalizations of $\pi_i$).

A subtlety concerns obtaining a proper expression for the variance of (17), on which the variance of the estimate hinges. First, due to the independence of different $\mathbf{P}_i$,

$$\text{Var} \sum_i \varphi_i(\mathbf{P}_i) = \sum_i \text{Var}\,\varphi_i(\mathbf{P}_i). \qquad (18)$$

Next, a predicament: how to estimate $\text{Var}\,\varphi_i(\mathbf{P}_i)$ if there is just one event $\mathbf{P}_i$ for each $i$? After a first shock, one realizes that the same reasoning would result in the same difficulty when all $\mathbf{P}_i$ are distributed identically, for which case a meaningful answer is nevertheless known: the mean value to subtract from $\varphi_i(\mathbf{P}_i)$ before squaring is the cumulative mean derived from the entire sample $\{\mathbf{P}_i\}_i$. This observation provides the needed clue. Indeed, the already obtained estimate $\theta_{\text{exp}}$ incorporates the cumulative information from the set $\{\mathbf{P}_i\}_i$. Then the required estimate for the mean is $h_i(\theta_{\text{exp}})$ (remember that all functions $h_i$ are assumed to be calculable). The right-hand side of Eq.(18) becomes:

$$\sum_i \big[\varphi_i(\mathbf{P}_i) - h_i\big(\theta_{\text{exp}}\big)\big]^2. \qquad (19)$$

Finally, one obtains an estimate for $\text{Var}\,\theta_{\text{exp}}$:

$$N\,\text{Var}\,\theta_{\text{exp}} = \frac{1}{NH^2}\sum_i \big[\varphi_i(\mathbf{P}_i) - h_i\big(\theta_{\text{exp}}\big)\big]^2. \qquad (20)$$

A consistency check is to set all $\pi_i$ equal. Then all weights $\varphi_i$, and all functions $h_i$ are equal too. Then the equation $h(\theta_{\text{exp}}) = \langle \varphi \rangle_{\text{exp}}$ is exactly the one used to obtain $\theta_{\text{exp}}$, and the familiar formula for the variance emerges. Which is хорошо.

Extension of the above formulae to the case of several parameters (sec. 3) is straightforward. One defines

$$\varphi_{\text{opt},i}^{(\alpha)}(\mathbf{P}) = \frac{\partial \ln \pi_i(\mathbf{P})}{\partial \theta_\alpha}, \qquad (21)$$

etc. The covariance matrix of the weights then is as follows (the independence of $\mathbf{P}_i$ is again important here):

$$\Phi_{\alpha,\beta} = \frac{1}{N}\sum_i \big[\varphi_i^{(\alpha)}(\mathbf{P}_i) - h_i^{(\alpha)}\big(\theta_{\text{exp}}\big)\big]$$
$$\times \big[\varphi_i^{(\beta)}(\mathbf{P}_i) - h_i^{(\beta)}\big(\theta_{\text{exp}}\big)\big]. \qquad (22)$$

Finally, the covariance matrix for the parameters $\theta$ is given by the relation (11) — but the determinant (13) can be evaluated directly from (22) and (12).

From here, one can proceed to programming.

## 5. OPTIMIZED LEAST SQUARES.

Instead of the structured iterations (7), with each $\theta$ a proper estimate obtained by solving the equations (17), one could choose to blindly minimize the "optimized" sum of squares (20) with respect to $\theta$ — one should then use the same $\theta$ in the construction of $\varphi_i$ (in place of $\theta_0$ etc.) and in the argument of $h_i$ in place of $\theta_{\text{exp}}$, i.e. in the probability distribution used to evaluate $h_i$ from $\varphi_i$ (see the example below). The resulting way to obtain the estimate is a generalization of the least squares method preserving, by construction, optimality in a general non-linear, non-gaussian case.

**NB** Unlike the venerable simplest sum of quarrés, the optimized sum of squares would be directly connected with the variance of the resulting estimate for $\theta$.

**NB** With several parameters to estimate, the determinant (13) replaces the sum (20) as a quantity to minimize.

**NB** An advantage of the iterative algorithm (7) over the blind minimization of the optimized sum of squares (20) (or the determinant (13)) is that one has a nice control point (with plots (4)) after each iteration because each $\theta_{\text{exp}(i)}$ is a meaningful estimate per se. (The plots are particularly useful if the weights are chosen such that the identity (8) holds.)

Those familiar with the character of multi-dimensional optimization would be reluctant to forgo such an advantage.

**6. Example.** Consider the problem of fitting a curve $y = f(x, \theta)$ with unknown $\theta$ against a set of points $\{x_i, y_i\}_i$, where $x_i$ are controlled by the experimenter and known precisely, whereas $y_i$ are distributed normally with the same variance $\sigma_i^2$ around the corresponding mean values $f(x_i, \theta)$. Then $\mathbf{P} \to y$, and the quasi-optimal weights can be chosen as $\varphi_{i,(0)}(y) = f'_\theta(x_i, \theta_0) y$, so that $h_i(\theta) = f'_\theta(x_i, \theta_0) f(x_i, \theta)$. To find $\theta_{\exp}$, one fits the theoretical number $N^{-1} \sum_i f'_\theta(x_i, \theta_0) f(x_i, \theta)$ against the experimental one, $N^{-1} \sum_i f'_\theta(x_i, \theta_0) y_i$. Then from (20):

$$N \operatorname{Var} \theta_{\exp} = H^{-2} N^{-1}$$
$$\times \sum_i [f'_\theta(x_i, \theta_0)]^2 [y_i - f(x_i, \theta_{\exp})]^2, \quad (23)$$

where

$$NH = \sum_i h'_i(\theta_{\exp}) = \sum_i f'_\theta(x_i, \theta_0) f'_\theta(x_i, \theta_{\exp}). \quad (24)$$

Then one can make a few iterations, using the newly found $\theta_{\exp}$ for $\theta_0$.

If one prefers to follow the optimized least squares route (sec. 5), then the expression to minimize would be the following sum:

$$\frac{\sum_i [f'_\theta(x_i, \theta)]^2 \times [y_i - f(x_i, \theta)]^2}{\left( \sum_i [f'_\theta(x_i, \theta)]^2 \right)^2}. \quad (25)$$

Its minimum value will be $\operatorname{Var} \theta_{\exp}$ (unlike the minimum value of the simplest sum of squares).

**NB** The new factors in the optimized sum of squares (25) are present despite the fact that the variance of $y$ is the same for all $x$.

Notice how the weights $[f'_\theta(x_i, \theta)]^2$ enhance (suppress) contributions from $x_i$ according to how fast (slow) $f(x_i, \theta)$ changes with varying $\theta$ — somewhat similarly to how the optimal weight (6) works.

To visualize this, consider examples where $\theta$ is a shift parameter, $f(x, \theta) = F(x - \theta)$, so that the steepness of the curve $y = F(x - \theta)$ directly reflects its sensitivity to changes in $\theta$. For example,

$$f(x, \theta) = \sin(x - \theta), \quad f'_\theta(x, \theta) = -\cos(x - \theta), \quad (26)$$

or a resonance curve

$$f(x, \theta) = [(x - \theta)^2 + \Gamma^2]^{-1},$$
$$f'_\theta(x, \theta) = 2(x - \theta)[(x - \theta)^2 + \Gamma^2]^{-2}. \quad (27)$$

**NB** The minimum of (25) corresponds to the Fisher-Fréchet-Rao-Cramér boundary, by construction.

**7. Conclusions.** The derived formulae set up a comprehensive successor to the least squares method, via a straightforward simple extension of the method of quasi-optimal weights [14], itself a systematic simple development of Pearson's long-neglected method of moments [2]. The new method inherits the nice analytical character of the classical method of moments — but also ensures asymptotic optimality of estimates. (An equivalence to the ML method was discussed in [14].) Unlike the ML method, it is insensitive to the details of shape of the probability density, and is thus applicable in some situations where the ML method fails due to the probability density having non-unique maxima or being only known approximately. The latter case may occur e.g. in perturbative calculations in quantum field theory, then evaluating weighted cross sections with smooth pre-agreed phase-space weights could significantly reduce the complexity of the task, bypassing the problem of singular non-positive higher-order corrections to matrix elements squared.

The method of quasi-optimal weights also leads, in a rather straightforward fashion (via the standard regularization trick), to other essential results of theoretical statistics such as the estimator $\frac{1}{2}(x_{\min} + x_{\max} - 1)$ for the $\theta$ of a uniform distribution on the segment $[\theta, 1 + \theta]$. More generally, the method remains meaningful with non-regular probability densities: then the FFRC minimum is simply located outside the space of continuous weights — but can be approached arbitrarily closely via regularization.

In the interesting case of experiments with rare decays where $\pi_i$ are Poisson distributions (e.g. [10]), the method has proved to be not in the least harder to implement than the standard weighted least squares method.

As for the potential improvement of physical results, a seasoned expert would not expect miracles. For example, Monte Carlo studies with simple models involving Poisson distributions show that a reduction of confidence intervals over the conventional least squares beyond 15% in the most opportune cases, is hardly likely. But even a 10% reduction of statistical errors is welcome as it would come essentially for free while being worth a 20% increase of data taking time, data storage, etc. In any event, such a gain proved to be compelling enough for [18], and an occasional bigger effect is possible [20]. Not to mention one thing less to worry about.

On the theoretical side, the most important result would not be any particular formula (although Eqs. (20) and (25) are cute). Rather, I believe, it would be the demonstration of economy and expediency of the functional-analytic interpretation of probability distributions. In a still broader context, this would be another argument to support the view (cf. also [7]) that although the interpretation of functions as correspondences $x \to y$ is an unassailable truth in the case of continuous functions, its naïve set-theoretic generalization, disrespectful of constructiveness, should give way in teaching and textbooks to the concept of generalized functions (distributions) as linear (perforce continuous [8], therefore constructive) functionals on proper test functions [5]. For "*the ultimate purpose which one should always keep in sight, is to find a correct point of view on the foundations of science*" (Karl Weierstrass). Perhaps instead of "foundations" one ought to be talking here about high-level abstractions that give the direction to our thoughts.

ACKNOWLEDGMENTS. I first heard least squares and optimal weights compared from the optimality viewpoint by Jörg Pretz [12], [18], while enjoying hospitality of the CERN theorists. V.M. Lobashev (of the Troitsk $\nu$-mass Experiment [10]) questioned adequacy of the least squares in a most specific context and kindly permitted the proposed answer to be tested on real data, in which undertaking the collaboration with A.A. Nozik, A.K. Skassyrskaya and S.V. Zadorozhny has been essential and enjoyable. V.Z. Nozik expressed an early appreciation of the idea. A.S. Barabash (of neutrinoless double beta-decay), A.V. Butkevich (of neutrino detection), V.K. Grigoriev (of hadron peakology) and Yu.G. Kudenko (of neutrino oscillations) showed a hearty interest in this work. Yu.M. Andreev (of supersymmetry), S.N. Gninenko (of solar axions) and Kh.S. Nirov (of quantum field theory) read parts of the manuscript. A.A. Borovkov (of mathematical statistics) looked through the first version of the text and supplied a piece of bibliographic information. G.A. Chernykh (of theoretical physics) provided a scan of Fock's trailblazing paper [5]. D. Sornette (of entrepreneurial risks) pointed out an adjacent body of work in the field of econometrics (e.g. [15]).

A.G. Dolgolenko and M.G. Kuzmina arranged presentations of the method before two sharp audiences — at the Institute of Theoretical and Experimental Physics and the Keldysh Institute of Applied Mathematics (both in Moscow). V.V.Vedenyapin, having independently arrived in [21] at the conclusion that the functional-analytic approach to probability *is better*, stood up for the cause.

I thank the two audiences and the INR theorists for their attention, and all the listed people for their understanding.

Support came from the CERN theory group, the RFBR grant 05-02-17238a, the Neutrino Physics Programme of the Russian Academy of Sciences, and the grant of the President of the Russian Federation NS-7293.2006.2 (RF government contract 02.445.11.7370).

## References
arranged chronologically